\documentclass{iaus}
\usepackage{graphicx}

\title[White Dwarf Model Atmospheres: Synthetic Spectra for Super Soft Sources] 
{White Dwarf Model Atmospheres:\\
Synthetic Spectra for Super Soft Sources}

\author[Thomas Rauch]   
{Thomas Rauch}

\affiliation{Institute for Astronomy and Astrophysics,
             Kepler Center for Astro and Particle Physics,
             Eberhard Karls University,
             Sand 1,
             72076 T\"ubingen,
             Germany\\email: {\tt rauch@astro.uni-tuebingen.de}}

\pubyear{2011}
\volume{281}  
\pagerange{119--126}
\jname{Binary Paths to the Explosions of Type Ia Supernovae }
\editors{R. Di Stefano \& M. Orio, eds.}
\begin{document}

\maketitle

\begin{abstract}
  The T\"ubingen NLTE Model-Atmosphere Package (\emph{TMAP}) 
calculates fully metal-line blanketed white dwarf model 
atmospheres and spectral energy distributions (SEDs) 
at a high level of sophistication. Such SEDs are easily
accessible via the German Astrophysical Virtual Observatory 
(\emph{GAVO}) service \emph{TheoSSA}.
We discuss applications of \emph{TMAP} models to (pre) white
dwarfs during the hottest stages of their stellar evolution,
e.g\@. in the parameter range of novae and super soft 
sources.
\end{abstract}

\firstsection 
\section{Introduction}
\label{sect:introduction}

Novae in outburst may become the brightest sources of soft X-ray emission
in the sky. During this so-called super soft source (SSS) phase
(some weeks or months after the outburst), 
it is possible to perform high-resolution and high-S/N spectroscopy with 
X-ray satellites like \emph{Chandra} or XMM-\emph{Newton}.
These spectra require adequate NLTE model atmospheres
for a reliable analysis.

\begin{figure}[t]
\begin{center}
 \includegraphics[width=\textwidth]{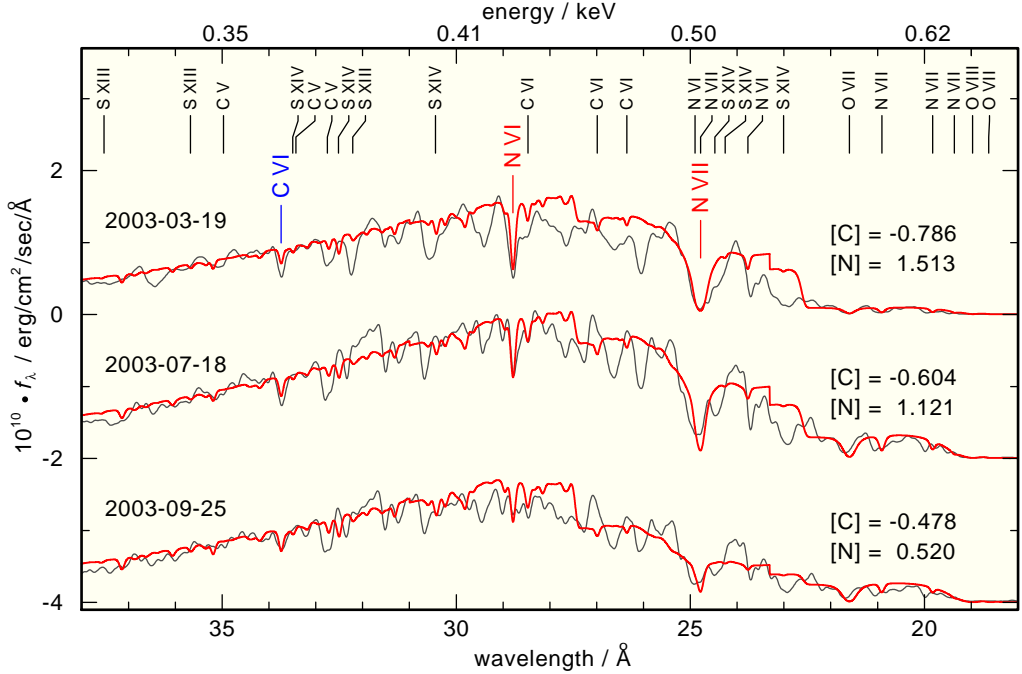}
 \caption{Comparison of flux-calibrated \emph{Chandra} observations of the nova V\,4743\,Sgr with
          synthetic spectra 
          (H+He+C+N+O+Ne+Mg+Si+S + (Ca$-$Ni) opacities considered in modeling, 
           $T_\mathrm{eff}\hspace{-0.5mm}=\hspace{-0.5mm}720\,000\,\mathrm{K}$, 
           $\log g = 9$, 
           cf\@. \cite[Rauch \etal\ 2010]{Rauch_etal_2010}). 
          Positions of line transitions are marked at top.
          $v_\mathrm{rad} = -2300\,\mathrm{km/sec}$ was applied to the models to match the
          C\,{\sc vi} and N\,{\sc vi} resonance lines.
          [X] denotes log\,(abundance / solar abundance).
          For clarity, the July and September observations are artificially shifted in flux.}
 \label{fig:cn}
\end{center}
\end{figure}

A detailed analysis of such a nova, V\,4743\,Sgr, was recently
presented by \cite{Rauch_etal_2010}. They used
plane-parallel, hydrostatic models
calculated with the T\"ubingen NLTE Model-Atmosphere Package 
(\cite[\emph{TMAP}\footnote{http://astro.uni-tuebingen.de/\raisebox{0.2em}{{\tiny $^\sim$}}TMAP/TMAP.html},
Werner \etal\ 2003, Rauch \& Deetjen\ 2003]
{Werner_etal_2003, RauchDeetjen_2003}).
Although the velocity field and the expansion of the nova's atmosphere was
neglected, the overall slope of the continuum flux was well reproduced.
Moreover, the strengths of prominent photospheric absorption lines
(C\,{\sc v}, C\,{\sc vi}, N\,{\sc vi}, N\,{\sc vii}, O\,{\sc vii})
as well as the strengths of absorption edges were in very good agreement with the observation
(Fig.\,\ref{fig:cn}, \ref{fig:final}). \cite{Rauch_etal_2010} could e.g\@.
show that the C/N abundance ratio was increasing from March to
September 2003 (Fig.\,\ref{fig:cn}).

\section{Model Atmospheres for SSSs}
\label{sect:model}

\emph{TMAP} model atmospheres were successfully employed for spectral analyses of hot, compact stars.
See \cite{RauchWerner_2010} and \cite{Rauch_2011} for a brief summary of this application to a variety of objects
from post-AGB stars ($T_\mathrm{eff}\hspace{-0.5mm}\approx\hspace{-0.5mm}100\,000\,\mathrm{K}$)
to neutron stars in
low-mass X-ray binaries ($T_\mathrm{eff}\hspace{-0.5mm}\approx\hspace{-0.5mm}10\,000\,000\,\mathrm{K}$).

{\it What do we need to analyze X-ray spectra taken during the SSS phase following a nova outburst reliably?}
The evolution of both stars has to be modeled, including e.g\@. stellar rotation, mass-loss history,
common-envelope phase, and a long series of outbursts. Thus, 3-dimensional 
magneto-hydrodynamical calculations that follow the stars for a long time are definitely the best choice.
In addition, 3D NLTE radiation transfer, accounting for the expanding atmosphere and
mass loss not only during the outburst, clumping in the ejected matter, etc\@. is necessary.
This should be combined with 3D photoionization modeling to describe the surrounding
gas and dust. Although all these state-of-the-art ingredients are already available,
their combination and application to an individual nova system is still a dream.

In the case of V\,4743 Sgr, \cite{Ness_etal_2003} measured a blueshift 
($v_\mathrm{rad} = - 2000\,\mathrm{km/sec}$) in the \emph{Chandra} spectrum.
This strong evidence for expansion was found in the SSS spectra of other novae as well (\cite[Ness 2010]{Ness_2010}).
A new version of the multi-purpose model-atmosphere code \emph{PHOENIX} 
(\cite[Hauschildt \& Baron 1999]{HauschildtBaron_1999})
accounts for this expansion (\cite[van Rossum \& Ness 2010]{vanRossumNess_2010}).
These models predict typical wind effects that are already known from
the UV wavelength range in cooler post-AGB stars. In the relevant parameter range for SSSs, the model flux 
for $\lambda < 30\,\mathrm{\AA}$ ($E > 0.41\,\mathrm{keV}$) is increasing with the mass-loss rate and absorption edges
of C\,{\sc vi} (25.30\,\AA), N\,{\sc vii} (18.59\,\AA), and O\,{\sc vii} (16.77\,\AA) become weaker.
From these results, one has to conclude that, of course only in case of higher mass-loss rates
($> 10^{-8}\,\mathrm{M_\odot / yr}$), spectral analyses of SSSs based on hydrostatic model atmospheres tend 
to overestimate $T_\mathrm{eff}$ and metal abundances.

\begin{figure}[t]
\begin{center}
 \includegraphics[width=\textwidth]{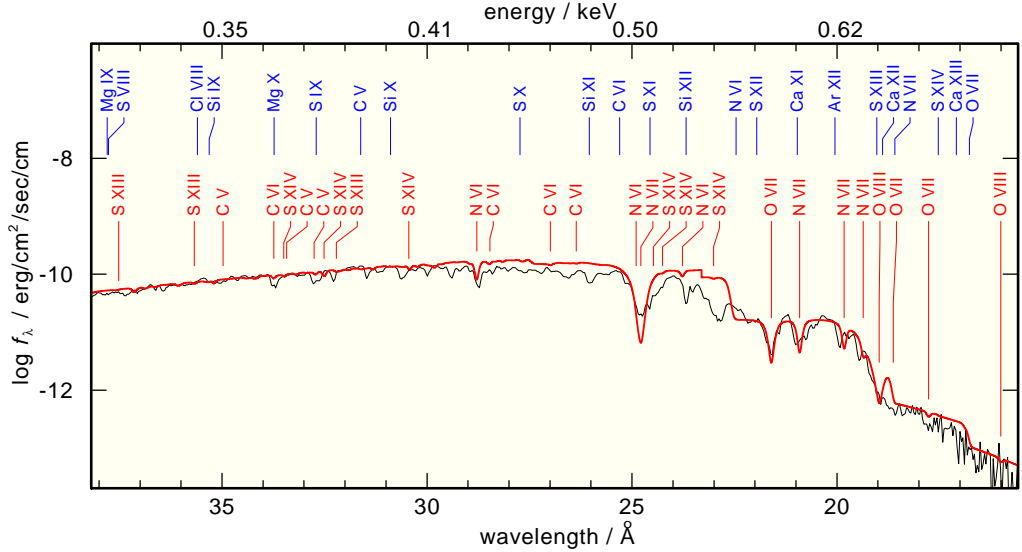} 
 \caption{Comparison of flux-calibrated XMM-\emph{Newton} observations of V\,4743\,Sgr (April 2003) with
          the final model 
          (H+He+C+N+O+Ne+Mg+Si+S + (Ca-Ni),
           $T_\mathrm{eff}\hspace{-0.5mm}=\hspace{-0.5mm}740\,000\,\mathrm{K}$, 
           $\log g = 9$)   
          of \cite{Rauch_etal_2010}.
          Positions of ground-state thresholds are marked at top (blue), those of line transitions
          just below that marks (red).} 
 \label{fig:final}
\end{center}
\end{figure}

The question remains, {\it why do the static TMAP models reproduce the observations that well} 
(Fig.\,\ref{fig:final}){\it?}
The answer may be very simple -- the strong mass-loss phase of V\,4743 Sgr is over. Obviously, there are no P\,Cygni line
profiles observed and thus, the ``wind-blown'' material around V\,4743 Sgr is not dense enough to
provide the emission. The matter in the line of sight is still moving towards us and thus, we see
blueshifted, pure absorption lines. An approach like presented by \cite{Rauch_etal_2010} appears
well justified.
It is possible that accretion already takes place and
changes the C/N abundance ratio (Sect.\,\ref{sect:introduction}).

\section{\emph{TheoSSA} - SSSs Model Spectra on Demand}
\label{sect:theossa}

\emph{TheoSSA}\footnote{Theoretical Stellar Spectra Access}
is a registered Virtual Observatory\footnote{http://www.ivoa.net} (\emph{VO}) service developed by the
German Astrophysical Virtual Observatory \emph{GAVO}\footnote{http://www.g-vo.org}.
It provides access to SEDs at three levels:
1) pre-calculated SED grids for a fast comparison with observations can be downloaded,
2) SEDs with individual parameters 
   (and provided standard model atoms from the
   T\"ubingen Model-Atom Database, 
   \emph{TMAD}\footnote{http://astro.uni-tuebingen.de/\raisebox{0.2em}{{\tiny $^\sim$}}TMAD/TMAD.html})
   can be calculated\footnote{http://astro.uni-tuebingen.de/\raisebox{0.2em}{{\tiny $^\sim$}}TMAW/TMAW.shtml} 
   without knowledge of the model-atmosphere code,
3) SEDs with individual parameters and self-constructed model atoms can be computed by experienced users.
   \emph{TheoSSA} was firstly based on \emph{TMAP} models only but it is prepared for SEDs from any other 
   stellar model-atmosphere code.

Pre-calculated grids of SEDs for SSSs (\cite[cf\@. Rauch \& Werner\ 2010]{RauchWerner_2010}) are also 
easily accessible in \emph{VO}-compliant 
form via \emph{TheoSSA}'s WWW interface\footnote{http://dc.g-vo.org/theossa}.
In case of any request, we will calculate new grids with different abundances.

In addition, some SED grids for SSSs 
($T_\mathrm{eff}\hspace{-0.5mm}=\hspace{-0.5mm}0.55 - 1.05\,\mathrm{MK}$, \cite[Rauch \& Werner 2010]{RauchWerner_2010}) 
are available converted into 
atables\footnote{http://astro.uni-tuebingen.de/\raisebox{0.2em}{{\tiny $^\sim$}}rauch/TMAF/flux\_HHeCNONeMgSiS\_gen.html}  
for the direct use within
\emph{XSPEC}\footnote{http://heasarc.gsfc.nasa.gov/docs/xanadu/xspec}.

\section{Conclusions}
\label{sect:conclusions}

Spectral energy distributions for SSSs (calculated with \emph{TMAP}) are easily available via \emph{TheoSSA} -- use them!
Blackbody energy distributions generally have -- at the same temperature -- 
a much lower peak intensity (about a factor of three in the parameter range of SSSs, \cite[Rauch \etal\ 2010]{Rauch_etal_2010})
and the flux maximum is located towards lower energies (about a factor of two).
Thus, ``spectral analyses'' based on blackbodies yield wrong results.

There are novae that are observed in their SSS phase where the impact of mass-loss is not significant.
For these objects, analyses based on static model atmospheres like \emph{TMAP} provide reliable results within
typical error ranges (\cite[Rauch \etal\ 2010]{Rauch_etal_2010}). In case of prominent P\,Cygni line profiles,
however, a code like \emph{PHOENIX} has to be employed. 

A time series (about all three months) of V\,4743 Sgr X-ray observations has shown that the 
white dwarf's surface is extremely hot ($T_\mathrm{eff}$ higher that 500\,000\,K)
for at least half a year (\cite[Rauch \etal\ 2010]{Rauch_etal_2010}). The phase of increasing $T_\mathrm{eff}$ appears
important because the mass-loss rate will be presumably higher and, hence,
expanding model atmospheres can be used to study wind properties. 

A time series of X-ray spectra on a short time scale (about weekly) 
starting at an early outburst phase and covering the SSS phase
like taken e.g\@. 
for Nova V\,1974 Cyg in the ultraviolet with IUE\footnote{International Ultraviolet Explorer} 
(cf\@. \cite[Gonz\'alez-Riestra \& Krautter 1998]{GonzalesKrautter_1998})
is highly desirable. Since \emph{Chandra} and XMM-\emph{Newton}
will unfortunately not work forever, such spectra have to be taken in the very near future.
The development of appropriate, reliable model atmospheres is already on the way.

\acknowledgments
This work is supported by 
the German Aerospace Center (DLR, grant 05\,OR\,0806),
the German Research Foundation (DFG, grant WE\,1312/41$-$1), and
the Federal Ministry of Education and Research (BMBF, grant 05A11VTB).

\end{document}